\documentclass[aps,pra,reprint,twocolumn,notitlepage,floatfix,nofootinbib,showkeys]{revtex4-2}

\usepackage{amsmath}
\usepackage{color}
\usepackage{graphicx}
\usepackage{hyperref}
\usepackage{multirow}
\usepackage{epigraph}

\newcommand{\bq}{\begin{eqnarray}}
\newcommand{\eq}{\end{eqnarray}}
\newcommand{\bqn}{\begin{eqnarray*}}
\newcommand{\eqn}{\end{eqnarray*}}
\newcommand{\bqs}{\begin{subequations}}
\newcommand{\eqs}{\end{subequations}}
\newcommand{\bw}{\begin{widetext}}
\newcommand{\ew}{\end{widetext}}
\newcommand{\xx}{{\bf x}}

\begin{document}
\title{Statistical Gravity and entropy of spacetime}

\author{Riccardo Fantoni}
\email{riccardo.fantoni@scuola.istruzione.it}
\affiliation{Universit\`a di Trieste, Dipartimento di Fisica, strada
  Costiera 11, 34151 Grignano (Trieste), Italy}

\date{\today}

\begin{abstract}
We discuss the foundations of the statistical gravity theory we proposed in a recent 
publication [Riccardo Fantoni, Quantum Reports, {\bf 6}, 706 (2024)].
\end{abstract}

\keywords{General Relativity; Einstein-Hilbert action; Statistical Physics; Path Integral; Entropy; Ergodicity}

\maketitle
\section{Introduction}
\label{sec:intro}

We propose a new horizontal theory which brings together statistical physics and general relativity.    

We give statistical physics \cite{LandauCTP5} foundation basis in order to determine the 
consistency of our theory, already put forward in Ref. \cite{Fantoni24f}, for a statistical 
gravity description. 

From a philosophycal point of view \cite{Ellis2006} we should ask about the mathematical issues of 
existence and unicity of the Universe as well as some anthropic questions like the fine tuning for 
life in our Universe or the natures of existence. We may think that before creation it was only chaos 
for which one could agree that between the two signatures of the metric of spacetime 
(the Euclidean and the Lorentzian) the one describing statistical physics (the Euclidean) would be 
the most appropriate. At creation, between before and after, it could be that one has to deal with
infinite energy densities or maybe density. From a description point of view we are already accustomed 
to deal with infinities. I am here thinking at the evolution of a Dirac delta into a Gaussian in a 
diffusion process. But there are many others.

The key logical point in the Theory we are proposing to explain the origins of gravity from the 
statistical approach, is the connection between thermodynamics and statistical physics made possible 
by the statistical concept of entropy and its derivative with respect to energy. This defines the 
temperature. In our statistical gravity theory the energy content is due to matter and 
electromagnetic fields and the entropy is a count of the quantum states of a quasi closed subregion 
of spacetime which can be considered closed for a period of time that is long relative to its 
relaxation time, with energy in a certain interval. Feynman will describe this in chapter 1 of his 
set of lectures \cite{FeynmanFIP} saying ``If a system is very weakly coupled to a heat bath at a 
given `temperature,' if the coupling is indefinite or not known precisely, if the coupling has been 
on for a long time, and if all the `fast' things have happened and all the `slow' things not, the 
system is said to be in {\sl thermal equilibrium}''. 

Our Eq. (\ref{eq:schroedinger}) has long been studied by John Klauder \cite{Fantoni23b} and the 
form chosen here is just representative and in substitution of the much more rigorous one offered 
by that author. Other alternative points of view are also present today 
\cite{Ashtekar86,Rovelli1993,Banerjee2010,Rovelli2013}.

This theory based on the mathematical properties of a Wick rotation would open a new sight of the 
statistical properties of spacetime as a physical entity.

Our theory can be considered a {\sl first step} towards a more sophisticated and dignified 
description of spacetime.

\section{Gentropy}

Let us define a {\sl subregion} of a macroscopic spacetime region as a part of spacetime that is 
very small respect to the whole Universe yet macroscopic.

The subregion is not closed. It interacts with the other parts of the Universe. Due to the large 
number of degrees of freedom of the other parts, the state of the subregion varies in a complex and 
intricate way.

In order to formulate a statistical theory of gravity we need to determine the {\sl statistical 
distribution} of a subregion of a macroscopic spacetime region. We know from General Relativity 
that each spacetime subregion has a metric so our statistical distribution will describe the 
statistical properties of these metric tensors $g_{\mu\nu}$.

Since different subregions ``interact'' weekly among themselves then:
\begin{itemize}
\item[1.] It is possible to consider them as {\sl statistically independent}, i.e. the state of a 
subregion does not affect the probability of the states of another subregion. If $\hat{\rho}_{12}$ 
is the density matrix of the subregion composed by the subregion 1 and by the subregion 2 then
\bq
\hat{\rho}_{12}=\hat{\rho}_1\hat{\rho}_2,
\eq
where $\hat{\rho}_i$ is the density matrix of the subregion $i$.
\footnote{General relativity is fundamentally a classical theory, while the density matrix is 
inherently quantum mechanical. This apparent contradiction will be solved in our discussion leading to 
Section \ref{sec:PI} when we will clarify which is the main actor that is in thermal equilibrium. As it 
will become clear then we think the metric tensor itself to be in thermodynamic equilibrium at a given 
temperature. Of course since the metric tensor determines the distances between events of the 
spacetime then this also implies that the spacetime itself is fluctuating due to thermal agitation.}
\item[2.] It is possible to consider a subregion as closed for a sufficiently small time interval. 
The time evolution of the density matrix of the subregion in such an interval of time is 
\bq \label{eq:schroedinger}
\frac{\partial}{\partial t} \hat{\rho}_i=
\frac{i}{\hbar}[\hat{\rho}_i,\hat{H}_i],
\eq
where $\hat{H}_i$ is the Hamiltonian of the quasi closed subregion $i$.

\item[3.] After a sufficiently long period of time the spacetime reaches the state of statistical 
equilibrium in which the density matrices of the subregions must be stationary. We must then have
\bq
[\prod_i\hat{\rho}_i,\hat{H}]=0,
\eq
where $\hat{H}$ is the Hamiltonian of the closed macroscopic spacetime. This condition is certainly 
satisfied if 
\bq
[\hat{\rho}_i,\hat{H}]=0,
\eq
for all $i$.
\end{itemize}

We then find that the logarithm of the density matrix of a subregion is an additive integral of 
motion of the spacetime.

This is certainly satisfied if 
\bq
\ln\hat{\rho}_i=\alpha_i+\beta_i \hat{H}_i.
\eq

In the time interval in which the subregion can be considered closed it is possible to diagonalize 
simultaneously $\hat{\rho}_i$ and $\hat{H}_i$. We then find
\bq \label{eq:alpha-beta}
\ln \rho_n^{(i)}=\alpha_i+\beta_i E_n^{(i)},
\eq
where the probabilities $\rho_n^{(i)}=w(E_n^{(i)})$ represent the distribution function in 
statistical gravity.

If we consider the closed spacetime as composed of many subregions and we neglect the 
``interactions'' among them, each state of the entire spacetime can be described specifying the 
state of the various subregions. Then the number $d\Gamma$ of quantum states of the closed spacetime 
corresponding to an infinitesimal interval of his energy must be the product
\bq
d\Gamma=\prod_i d\Gamma_i,
\eq
of the numbers $d\Gamma_i$ of the quantum states of the various subregions.

In fact we will have an uncertainty principle \cite{Karolyhazy1966,Maziashvili2007} ruling the two 
conjugated variables that are the 
generalized `coordinate': The metric tensor field $g_{\mu\nu}(x)$ where $x$ is an event of spacetime 
and the generalized `momentum': The operator $\hat{\pi}^{\mu\nu}=-i\hbar\delta/\delta g_{\mu\nu}$. As 
usual we will have
\bq \nonumber
\Delta g_{\mu\nu}(x)\Delta\hat{\pi}^{\alpha\beta}(x')&\ge&\frac{1}{2}
|\langle[g_{\mu\nu}(x),\hat{\pi}^{\alpha\beta}(x')]\rangle|\\
&=&\frac{1}{2}\hbar\delta^{(4)}(x-x')\delta_\mu^\alpha\delta_\nu^\beta,
\eq
where $\langle\ldots\rangle$ denotes a vacuum expectation value, $\Delta$ indicates a standard deviation, 
$\delta^{(4)}$ is a Dirac delta function in 4 dimensions, and the other two $\delta$ are Kronecker 
symbols. 
\footnote{It has been pointed out by prof. John R. Klauder that in the Arnowitt, Deser, and Misner 
\cite{Fantoni24l} $3+1$ foliation scheme, that seems to be necessary to treat the path integral 
described in Section \ref{sec:PI}, where the imaginary time naturally splits from the space, the 
generalized `coordinate' is played by the spatial components of the metric tensor field. But these
subtensor must be positive definite. So that due to this anholonomous constraint the corresponding
generalized `momentum' would cease to be a self-adjoint operator. In order to put things back in 
order the most elegant way is to use {\sl Affine Quentization} that amounts to define a different
generalized momentum, the so called {\sl dilation} operator $\hat{\pi}_a^b=g_{ac}\hat{\pi}^{cb}$, 
where we indicate with a latin index a spatial component. So doing the dilation operator is made
self-adjoint by construction. \label{foot:aq}}
Here we are associating $g_{\mu\nu}(x)$ with the metric tensor from General Relativity entering 
our Eq. (\ref{eq:ltdm}). A delicate point is that of a consistent description of the vacuum of 
General Relativity where both matter fields and the Ricci scalar vanish for which our high 
temperature density matrix of Eq. (\ref{eq:ltdm}) reduces to a functional Dirac delta.

We can then formulate the expression for the {\sl microcanonical distribution function} writing
\bq
dw\propto\delta(E-E_0)\prod_i d\Gamma_i
\eq
for the probability to find the closed spacetime in any of the states $d\Gamma$.

Let us consider a spacetime that is closed for a period of time that is long relative to its 
relaxation time. This implies that the spacetime is in complete statistical equilibrium. 

Let us divide the spacetime region in a large number of macroscopic parts and consider one of these. 
Let $\rho_n=w(E_n)$ be the distribution function for such part. In order to obtain the probability 
$W(E)dE$ that the subregion has an energy between $E$ and $E+dE$ we must multiply $w(E)$ by the 
number of quantum states with energies in this interval. Let us call $\Gamma(E)$ the number of 
quantum states with energies less or equal to $E$. Then the required number of quantum states with 
energy between $E$ and $E+dE$ is
\bq
\frac{d\Gamma(E)}{dE}dE,
\eq
and the energy probability distribution is 
\bq
W(E)=\frac{d\Gamma(E)}{dE}w(E),
\eq
with the normalization condition
\bq
\int W(E)dE=1.
\eq

The function $W(E)$ has a well defined maximum in $E=\bar{E}$. We can define the ``width'' 
$\Delta E$ of the curve $W=W(E)$ through the relation 
\bq
W(\bar{E})\Delta E=1.
\eq
or
\bq
w(\bar{E})\Delta\Gamma=1,
\eq
where
\bq
\Delta\Gamma=\frac{d\Gamma(\bar{E})}{dE}\Delta E,
\eq
is the number of quantum states corresponding to the energy interval $\Delta E$ at $\bar{E}$. 
This is also called the {\sl statistical weight} of the macroscopic state of the subregion, and its 
logarithm
\bq
S=\log\Delta\Gamma,
\eq
is the {\sl entropy} of the subregion. The entropy cannot be negative.

We can also write the definition of entropy in another form, expressing it directly in terms of the 
distribution function. In fact we can rewrite Eq. (\ref{eq:alpha-beta}) as
\bq
\log w(\bar{E})=\alpha+\beta\bar{E},
\eq
so that
\bq \nonumber
S=\log\Delta\Gamma&=&-\log w(\bar{E})=-\langle\log w(E_n)\rangle\\
&=&-\sum_n\rho_n\log\rho_n=-\mbox{tr}(\hat{\rho}\log\hat{\rho}),
\eq
where `tr' denotes the trace. 

Let us now consider again the closed region and let us suppose that 
$\Delta\Gamma_1,\Delta\Gamma_2, \ldots$ are the statistical weights of the various subregions,
then the statistical weight of the entire region can be written as 
\bq
\Delta\Gamma=\prod_i\Delta\Gamma_i,
\eq
and
\bq
S=\sum_iS_i,
\eq
the entropy is additive.

Let us consider again the microcanonical distribution function for a closed region,
\bq \nonumber
dw&\propto&\delta(E-E_0)\prod_i\frac{d\Gamma_i}{dE_i}dE_i\\
\nonumber
&\propto&\delta(E-E_0)e^S\prod_i\frac{dE_i}{\Delta E_i}\\
&\propto&\delta(E-E_0)e^S\prod_i dE_i,
\eq
where $S=\sum_iS_i(E_i)$ and $E=\sum_iE_i$. Now we know that the most probable values of the 
energies $E_i$ are the mean values $\bar{E_i}$. This means that the function $S(E_1,E_2,\ldots)$ 
must have its maximum when $E_i=\bar{E_i}$ for all $i$. But the $\bar{E_i}$ are the values of the 
energies of 
the subregions that correspond to the complete statistical equilibrium of the region. We then reach 
the important conclusion that the entropy of a closed region in a state of complete statistical 
equilibrium has its maximum value (for a given energy of the region $E_0$).

Let us now consider again the problem to find the distribution function of the subregion, i.e. of 
any macroscopic region being a small part of a large closed region. We then apply the microcanonical 
distribution function to the entire region. We will call the ``medium'' what remains of the 
spacetime region once the small macroscopic part has been removed. The microcanonical distribution 
can be written as
\bq
dw\propto\delta(E+E^\prime-E_0)d\Gamma d\Gamma^\prime,
\eq
where $E,d\Gamma$ and $E^\prime,d\Gamma^\prime$ refer to the subregion and to the ``medium'' 
respectively, and $E_0$ is the energy of the closed region that must equal the sum $E+E^\prime$
of the energies of the subregion and of the medium.

We are looking for the probability $w_n$ of one state of the region so that the subregion is in some 
well defined quantum state (with energy $E_n$), i.e. a well defined microscopic state. Let us then 
take $d\Gamma=1$, set $E=E_n$ and integrate respect to $\Gamma^\prime$
\bq \nonumber
\rho_n&\propto&\int\delta(E_n+E^\prime-E_0)d\Gamma^\prime\\
\nonumber
&\propto&\int\frac{e^{S^\prime}}{\Delta E^\prime}
\delta(E_n+E^\prime-E_0)dE^\prime\\
&\propto&\left(\frac{e^{S^\prime}}{\Delta E^\prime}
\right)_{E^\prime=E_0-E_n}.
\eq

We use now the fact that, since the subregion is small, its energy $E_n$ will be small respect to 
$E_0$
\bq
S^\prime(E_0-E_n)\approx S^\prime(E_0)-E_n\frac{dS^\prime(E_0)}{dE_0}.
\eq
But we know that the derivative of the entropy with respect to the energy is $\beta=1/k_BT$ where 
$k_B$ is Boltzmann constant and $T$ is the temperature of the closed spacetime region (that 
coincides with that of the subregion with which it is in equilibrium). So we finally reach the 
following result
\bq
\rho_n\propto e^{-\beta E_n}.
\eq
which is the {\sl canonical distribution function}. 

\section{Metric representation of the density matrix and path integral}
\label{sec:PI}

We then reach to the following expression for the density matrix of spacetime
\bq
\hat{\rho}\propto e^{-\beta\hat{H}},
\eq
where $\hat{H}$ is the spacetime Hamiltonian. In the non-quantum high temperature regime we can let 
$\beta\to\beta/M$ with $M$ a large integer. Then we can use for the high temperature density matrix
the usual classical limit \cite{Fantoni24f,Fantoni24l,Schulman,Ceperley1995}
\begin{widetext}
\bq \label{eq:ltdm}
\rho(\{g_{\mu\nu}\},\{g'_{\mu\nu}\};\tau)\propto\exp\left[-\tau\int_\Omega \left(\frac{1}{2\kappa}R+{\cal L}_F\right)\, \sqrt{g}\,d^3\xx\right],
\eq
\end{widetext}
where $g_{\mu\nu}(x)$ is the spacetime metric tensor, $x\equiv (ct,\xx)=(x^0,x^1,x^2,x^3)$ is an 
event in space($\xx$)time($t$), $\tau=\beta/M$ is a small complex time step, 
$R$ is the Ricci scalar of the spacetime subregion, $\kappa =8\pi Gc^{-4}$ is Einstein's 
gravitational constant ($G$ is the gravitational constant and $c$ is the speed of light in vacuum), 
$\Omega$ is the volume of space of the subregion whose spacetime is curved by the matter and 
electromagnetic fields due to the term ${\cal L}_F$, and $g$ is the determinant of the metric tensor. 

Using then Trotter formula \cite{Trotter1959} we reach to the path integral expression described in 
Ref. \cite{Fantoni24f} for the finite temperature case, where the metric tensor path wanders in the 
spacetime subregion made of the complex time interval $[0,\hbar\beta[$ with periodic boundary 
conditions and the spatial region $\Omega$. The spatial region can be compact in the absence of 
black holes or not if any are present. In any case it can either include its outermost frontier or 
not but from a numerical point of view it is convenient to use periodic boundary conditions there in 
order to simulate a thermodynamic limit so that only the frontiers around eventual black holes 
matter. The metric tensor 10-dimensional space is an hypertorus with 
$g_{\mu\nu}(ct+\beta(\xx),\xx)=g_{\mu\nu}(ct,\xx)$ and 
$g_{\mu\nu}(ct,\xx+\boldsymbol{\xi})=g_{\mu\nu}(ct,\xx)$. In the classical regime, when $\beta$ 
is small, and if the periodicities along different spatial dimensions are incommensurable, i.e. 
$\xi^i/\xi^j$ for $i\neq j$ cannot be written as rational numbers, then the Einstein field 
equations will let the metric tensor explore its phase space in a quasi-periodic fashion, 
so that one can use either a ``molecular-'' (or ``hydro-'') dynamics numerical simulation strategy, 
since the imaginary time averages equals the ensemble averages thanks to ergodicity, or a Monte Carlo 
numerical simulation strategy. In the quantum regime, when $\beta$ is big, it is necessary to
use the Path Integral Monte Carlo method described above. 

The field theory we are approaching with our path integral method where the main actor is the 
metric tensor field may be subject to triviality problems as the ones that occur for example 
when treating the $\varphi^r_n$ scalar euclidean covariant relativistic quantum field theory 
\cite{Fantoni23b} for $r\ge 2n/(n-2)$ where $n$ is the number of spacetime dimensions and $r$ the 
integer positive power of the interaction term $g|\varphi(x)|^r$, $g$ being the coupling constant.
It could be that as happen in that scalar case also for the tensorial case studied here, affine 
quantization plays a crucial role. As already mentioned in footnote \ref{foot:aq} affine 
quantization is a necessary tool to treat a constrained theory with mathematical rigor. 
Moreover it is different from the quantization proposed by Ashtekar \cite{Ashtekar86} and
as such it is novel. 

\section{Conclusions}

We gave logical foundation to the statistical gravity horizontal theory we recently 
proposed \cite{Fantoni24f,Fantoni24l}. Our weakness in discussing Eq. (\ref{eq:schroedinger}) does 
not reflect a weakness in the current knowledge and studies around that equation but is just our 
lack of deep vertical awareness. 

Our statistical theory of gravitation defines a 
temperature of the metric tensor which measures the distances between events in spacetime as the
derivative of the energy of the metric respect to its entropy.

\section*{Author declarations}
\subsection*{Conflict of interest}
The author has no conflicts to disclose.

\section*{Data availability}
The data that support the findings of this study are available from the 
corresponding author upon reasonable request.
\bibliography{gentropy}

\end{document}